\documentclass[runningheads]{llncs}

\usepackage[T1]{fontenc}
\usepackage{newtxtext}
\usepackage{amsmath}
\usepackage[varvw]{newtxmath}

\usepackage{graphicx}
\usepackage{listings}
\usepackage{xcolor}
\usepackage{booktabs}
\usepackage{tabularx}
\usepackage{placeins}
\usepackage{needspace}
\usepackage{hyperref}

\hypersetup{
  pdftitle={Qaiji-IR: An Eight-Layer Intermediate Representation Family for Hybrid Quantum-Classical Compilation},
  pdfauthor={Ye Jun},
  pdfsubject={PDCAT 2026, Track 11},
  pdfkeywords={Hybrid quantum-classical compilation, Intermediate representation, Semantic preservation, Semantic contract, Compiler architecture}
}

\lstdefinestyle{pythonsnippet}{
  basicstyle=\ttfamily\footnotesize,
  keywordstyle=\bfseries,
  stringstyle=\normalfont\ttfamily,
  commentstyle=\itshape,
  language=Python,
  showstringspaces=false,
  frame=single,
  framerule=0.3pt,
  rulecolor=\color{gray!50},
  numbers=none,
  breaklines=true,
}

\newcommand{\projectname}{\texorpdfstring{Qaiji-IR}{Qaiji-IR}}
\begin{document}
\renewcommand{\thelstlisting}{\arabic{lstlisting}}


\title{\projectname: An Eight-Layer Intermediate Representation Family for Hybrid Quantum-Classical Compilation}
\titlerunning{\projectname}

\author{Ye Jun\inst{1,2}\orcidID{0000-0003-1963-0865}}
\authorrunning{Ye Jun}
\institute{CETC International Cornerstone Quantum Industry (Suzhou) Co., Ltd., Suzhou, China \\
  \email{yjmaxpayne@hotmail.com}
  \and
  Institute for Quantum Science, School of Physical Science and Technology, Soochow University, Suzhou 215006, China}

\maketitle


\begin{abstract}
Hybrid quantum-classical compilation relates ideal operations and measurement outcomes to control pulses and device instructions. Existing formats each describe part of this process, and their preservation checks are often specific to one tool rather than to a common intermediate representation (IR). I present \projectname, an eight-layer IR family organized by a \emph{five-axis semantic contract} for data, semantics, lowering, runtime, and verification. Its implementation provides typed circuits, an operation-semantics layer (L4) that records operation structure and checks a bounded set of equivalence rules, and a native-scheduling layer (L2) with separate validation.

Nine circuit pairs are compared with ideal reference calculations. An aggregate baseline, which compares operation categories and measurement counts, accepts four behavior-changing transformations: gate reordering and changes of rotation angle, operand, or measurement destination. The L4 checker passes none of them. It distinguishes strict equality from equality up to global phase and returns unsupported when its rules cannot decide, as for valid rotation fusion. L2 validation checks native schedules against their source circuits, decomposition rules, local matrices, and configured timing.

A single-qubit active-reset program is compiled through the Qaiji-IR frontend, a syntax adapter, and the EmuPlat backend into pulses and device instructions for a virtual pulse processing unit (VPPU). Injected binary outcomes select the conditional drive. Backend predicate tests reveal incorrect execution for equality to zero and for a nonbinary feedback value. This backend path does not use the validated L2 schedule. Physical measurement feedback, calibrated pulse fidelity, and feedback latency require further work.

\end{abstract}

\section{Introduction}
\label{sec:intro}

Consider a small measurement-conditioned program: apply $H$ to qubit $q_1$, apply a CNOT with control $q_0$ and target $q_1$, apply $R_Z(\pi/2)$ to $q_0$, measure $q_1$ into classical bit $c_0$, and conditionally apply a gate to $q_0$. Even this five-operation fragment crosses several compiler boundaries. The circuit representation contains gates and a measurement; the hybrid-control representation carries the dependency from the measurement result to the branch; pulse and device representations supply timing and control instructions; and physical execution depends on calibration, the Hamiltonian, and the noise model. Assessing preservation across these representations requires separating agreement of operation labels from agreement of the operators and measurement records.

Existing representations make different, useful boundary choices. OpenQASM~3 includes classical control, timing, and embedded pulse-level calibration definitions~\cite{openqasm3}; QIR represents quantum programs within LLVM IR~\cite{qir2021}; and MLIR-based systems such as CUDA-Q, QLLVM, and Catalyst provide extensible dialect and lowering infrastructure~\cite{cudaq,qllvm,catalyst}. IBM qe-compiler exposes multiple abstraction levels, including pulse operations~\cite{ibm-qe-pulse}, while QubiC defines a control stack and instruction path for mid-circuit feed-forward~\cite{qubic2}. \projectname{} makes operation structure and bounded equivalence checks explicit; an experiment with the EmuPlat backend~\cite{emuplat} examines the connection to pulse and instruction-set architecture (ISA) lowering.

Relating quantum operations and measurement records to scheduled controls requires a data model, semantics, lowering rules, runtime behavior, and verification methods. I call this a \emph{five-axis semantic contract}. The name \projectname{} alludes to the interaction of water and fire in shaping clay in Song Yingxing's \emph{Tian Gong Kai Wu}~\cite{song1637}; here it refers to combining quantum states, classical control, pulse signals, and physical constraints within one compilation model. \projectname{} organizes the contract as an eight-layer IR family. Its L4 SemanticIR stores operation classes, declared equivalence levels, optimization hints, and classical-feedback models as attributes.

The paper makes three contributions. First, it defines the five-axis contract across six main layers plus dynamics and trace side layers. Second, the evaluated Qaiji-IR implementation provides typed circuits, L4 rule checks, and separately validated L2 schedules. Third, semantic reference calculations and a single-source backend experiment examine their scope, including VPPU execution of a binary active-reset branch.

The two-qubit fragment illustrates the upper representation layers. For execution, I use a one-qubit active-reset source that passes through the Qaiji-IR frontend, canonical QASM, a syntax adapter, and the EmuPlat backend. Qaiji-IR's L4/L2 checks are evaluated directly. The backend uses its own annotations and aggregate check, and its pulse/ISA path does not consume the validated native schedule.

\S\ref{sec:background} introduces the four concerns and five-axis contract. \S\ref{sec:arch} describes the IR family, its checks, and the backend pipeline. \S\ref{sec:impl} follows the example to execution, and \S\ref{sec:eval} evaluates the checks and predicate support. \S\ref{sec:related} compares related IR systems, and \S\ref{sec:conclusion} concludes.

\section{Background}
\label{sec:background}

\subsection{Four Concerns in Hybrid Quantum-Classical Compilation}
\label{sec:four-logics}

A hybrid quantum-classical compiler must coordinate four concerns. \emph{Application} semantics describes algorithms, workflows, logical gates, and resource objectives. \emph{Information} semantics describes typed quantum and classical dataflow, including measurement results, branch dependencies, and Pauli-frame or register handles. \emph{Signal} semantics maps native operations to scheduled pulses, waveforms, and hardware instructions. \emph{Physical} semantics relates those control artifacts to Hamiltonians, noise and leakage models, and calibration state. These four concerns provide the organizing model for \projectname.

Operator equality, measurement-result labeling, and pulse timing require different checks. Equality of ideal circuit operators alone does not establish calibrated pulse fidelity. Existing toolchains distribute these checks across language specifications, dialects, compiler passes, runtime interfaces, and hardware descriptions. \projectname{} brings these responsibilities into a common compilation model.

\subsection{The Five-Axis Semantic Contract}
\label{sec:five-axis}

\begin{figure}[htbp]
\centering
\includegraphics[width=\linewidth]{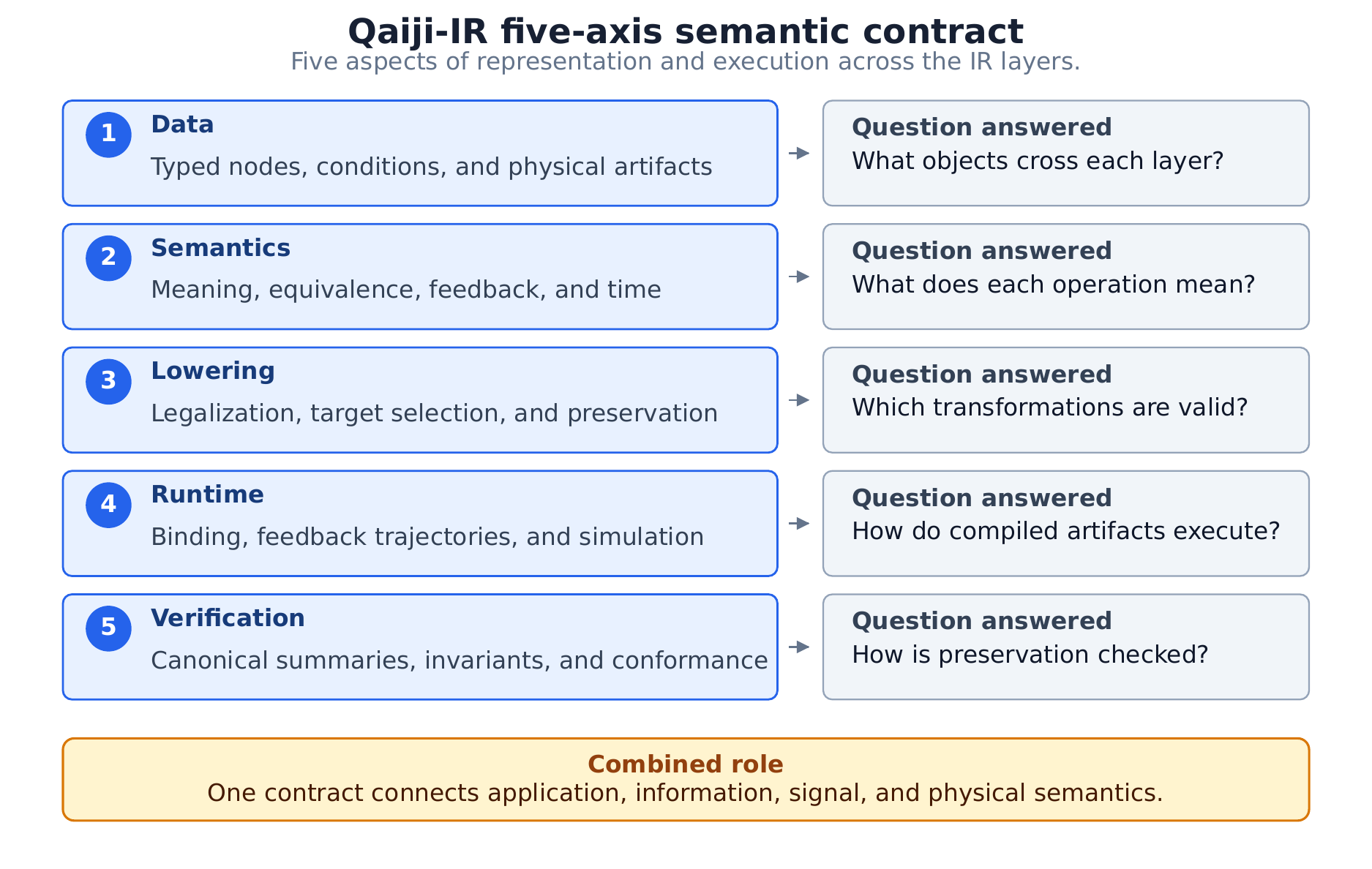}
\Description{Five rows describe the Qaiji-IR semantic contract. Data defines the objects passed across layers. Semantics defines operation meaning, equivalence, feedback, and time. Lowering describes legalization, target selection, and preservation. Runtime connects binding, feedback, and simulation to execution. Verification describes checks based on summaries, invariants, and conformance. Together these connect application, information, signal, and physical semantics.}
\caption{The five-axis semantic contract connects the four concerns in \S\ref{sec:four-logics}. Each row is one axis.}
\label{fig:contract}
\end{figure}

The semantic contract has five parts. Data and semantic definitions specify the IR objects and their meaning, including measurement handles, feedback edges, pulses, and dynamics. Lowering rules describe legalization, target selection, and the preservation information passed between stages. Runtime requirements connect compiled programs to execution and physical simulation. Verification covers canonical summaries, invariants, preservation checks, and conformance across implementations. Figure~\ref{fig:contract} pairs each part with the question it answers; \S\ref{sec:arch} distributes these responsibilities across eight layers.

\subsection{Separating Semantic Types from Device Opcodes}
\label{sec:sovereignty-in-code}

Operation meaning and device opcodes require separate interfaces. Qaiji-IR L4 records semantics without ISA types; its L2 projection validates a native schedule before emitting neutral operation labels. In the EmuPlat backend, the mapping in Listing~\ref{lst:qubic-mapping} assigns action labels to semantic categories. Measurement requests readout and result registration; conditional control requests arithmetic and a jump. The backend adapter then selects concrete opcodes. These backend action labels are not part of the Qaiji-IR L4 API.

\noindent\begin{minipage}{\linewidth}
\begin{lstlisting}[style=pythonsnippet,
  caption={EmuPlat backend mapping from semantic tags to ISA-neutral action labels.},
  label={lst:qubic-mapping}]
_DEFAULT_MAPPING: dict[MorphismType, tuple[str, ...]] = {
    MorphismType.UNITARY:        ("PULSE",),
    MorphismType.PERMUTATION:    ("PULSE",),
    MorphismType.MEASUREMENT:    ("PULSE_READOUT", "REGISTER_RESULT"),
    MorphismType.PHASE:          ("VIRTUAL_PHASE",),
    MorphismType.IDENTITY:       ("IDLE",),
    MorphismType.CLASSICAL_CTRL: ("ALU", "JUMP"),
}
\end{lstlisting}
\end{minipage}

\FloatBarrier
\section{\projectname{} Architecture}
\label{sec:arch}
\subsection{The Eight-Layer IR Family}
\label{sec:layers}
Six main layers describe program structure (L5), operation semantics (L4), hybrid control (L3), native scheduling (L2), pulses (L1), and sampled waveforms (L0). Dynamics P and traces T are side layers. Throughout this paper, the evaluated Qaiji-IR implementation is main-branch commit \texttt{4ef6f1a}, verified on 2026-10-18~\cite{qaiji-ir}. Table~\ref{tab:coverage} separates its public interfaces from external backend components. Qaiji-IR implements L5 program records, typed circuits, L4 semantics, and L2 scheduling. L3 dependencies occur in conditional nodes and measurement references; there is no general control-flow graph. L1/L0/P/T are architectural roles outside the package's public API. EmuPlat supplies the pulse/ISA/VPPU path for the experiment described below.

\begin{table}[htbp]

\caption{Eight architectural roles: implementing Qaiji-IR or external backend component, evaluated functionality, and further work.}
\label{tab:coverage}
\scriptsize
\begin{tabularx}{\linewidth}{@{}l>{\raggedright\arraybackslash}X>{\raggedright\arraybackslash}X>{\raggedright\arraybackslash}X@{}}
\toprule
Role & Implementation & Evaluated functionality & Further work \\
\midrule
L5 & Qaiji-IR program records and typed Circuit nodes & Codec roundtrip; program regression tests & Wider language coverage \\
L4 & Qaiji-IR annotations, structural summaries, R1--R4 checks & Nine pairs against ideal reference & Broader transformations \\
L3 & Conditional nodes and measurement dependencies & One condition; L2 readiness checks & General predicates and CFG \\
L2 & Qaiji-IR CZ-basis schedules and separate validators & Source, rule, matrix, and timing probes & Connect schedule to backend \\
L1 & External EmuPlat symbolic pulse templates & Explicit template and ISA binding & Calibrated pulse equivalence \\
L0 & External VPPU waveform facilities & Encoded program yields drive samples & DAC/RTL or hardware equivalence \\
P & External dynamics models and solver APIs & Dynamics not run in this example & Collapse then conditional continuation \\
T & Backend/driver logs and certificate metadata & Recorded instructions and branch PCs & Unified traces and proof production \\
\bottomrule
\end{tabularx}
\end{table}

\begin{figure}[htbp]
\centering
\includegraphics[width=.95\linewidth]{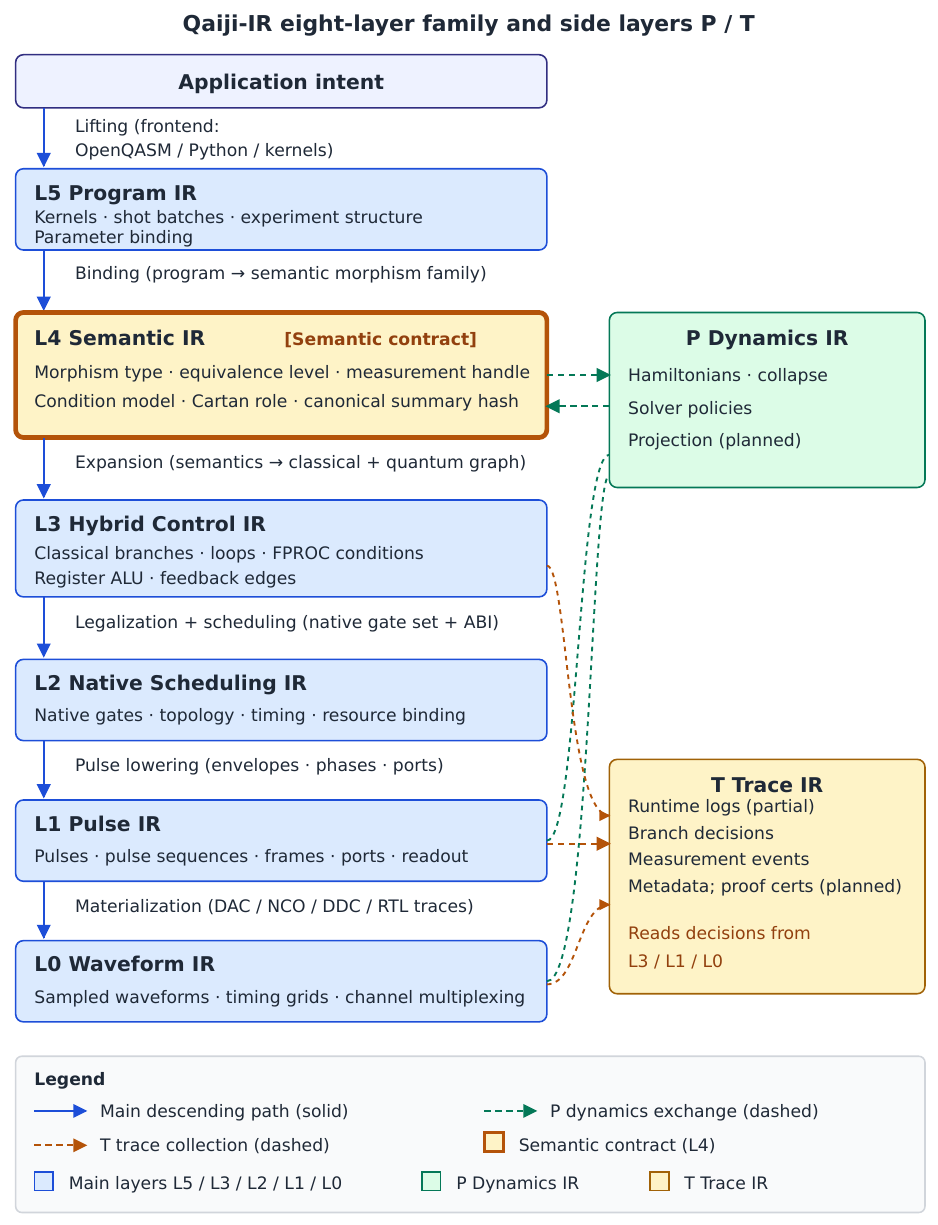}
\Description{Eight architectural layers. L5 to L0 are the main roles; P dynamics and T traces are side roles. Projection and automatic preservation certificates are marked as goals; certificate metadata is implemented. Table 1 describes the implementation and evaluated functions.}
\caption{Architectural topology, not package coverage. Table~\ref{tab:coverage} identifies the Qaiji-IR and backend interfaces. Dashed links, dynamics projection, and automatic preservation certificates denote further work.}
\label{fig:topology}
\end{figure}

\subsection{Operation Semantics in L4}
\label{sec:l4-semantic}
Table~\ref{tab:schema} lists the Qaiji-IR records. An annotation's \texttt{gate\_index} identifies a top-level node; \texttt{body\_offset} identifies a gate inside a conditional body. Measurement handles retain the register name and bit index. Classical edges declare register dependencies, but do not prove causal order; L2 separately checks preceding measurements and configured readiness. The summary includes ordered operations, operands, parameters, measurement destinations, and condition bodies.

\begin{table}[htbp]

\caption{Core records in the evaluated Qaiji-IR main commit.}
\label{tab:schema}
\scriptsize
\begin{tabularx}{\linewidth}{@{}l>{\raggedright\arraybackslash}X@{}}
\toprule
Record & Fields and interpretation \\
\midrule
Annotation & \texttt{gate\_index}, \texttt{body\_offset}, \texttt{morphism}, \texttt{equiv\_level}, \texttt{cartan}, \texttt{notes}. Conditional node and body gates have separate entries. \\
Measurement handle & \texttt{measurement\_id}, \texttt{qubit\_index}, \texttt{creg\_name}, \texttt{creg\_index}, \texttt{source}. \\
Classical edge & \texttt{source\_handle}, \texttt{target}, \texttt{edge\_kind}; a declarative measurement-to-condition dependency. \\
Preservation & \texttt{stage}, \texttt{status}, \texttt{diagnostic\_count}, \texttt{failed\_gate\_indices}, \texttt{equiv\_level}, \texttt{original\_morphism}, \texttt{lowered\_morphism}, \texttt{pass\_trace\_ref}. \\
L4 summary & Schema \texttt{qaiji.semantic\_summary.v0}; \texttt{num\_qubits}, \texttt{cregs}, ordered \texttt{operations}, annotation counts, handles, edges, and \texttt{condition\_model}. Hash identifies structure. \\
Native schedule & \texttt{source\_kernel\_ref}, \texttt{operations}, \texttt{groups}, \texttt{events}, \texttt{conditions}, \texttt{end\_ns}; also qubit/register and schema/ruleset versions. \\
Native validation & Separate \texttt{source}, \texttt{rules}, \texttt{hp} (local matrices), \texttt{schedule}, per-group evidence, and \texttt{all\_results\_ready\_ns}. \\
\bottomrule
\end{tabularx}
\end{table}

An operation receives a \textsf{MorphismType} tag, but the corresponding mathematical classes can overlap: permutations and phase gates are also unitary. The registry assigns $I$ to identity, $X$/CNOT/SWAP to permutation, $Z/S/T/CZ$ to phase, and $H/Y/R_X/R_Y$ to unitary. $R_Z(\theta)$ is phase except at numerically recognized full turns, where it receives the identity tag. A conditional node is classical control; its body gates are annotated separately with their own categories. Cartan roles are advisory optimization hints, not inputs to the equivalence decision.

For unitaries, strict equality means $U=V$; equality up to global phase means $U=e^{i\phi}V$. For measurements, equality compares the quantum operation for each classical result label. Qaiji-IR checks position-aligned rules: structural agreement (R1), CNOT/CX aliasing (R2), same-axis rotation differences of $2\pi k$ (R3), and full-turn rotation/identity folds (R4). Gate pairs must have identical operands. Measurement destinations and condition register, value, and body structure participate in the comparison. Numerical angle comparisons use absolute tolerance $10^{-10}$.

The rules assign \textsf{EXACT} or \textsf{UP\_TO\_PHASE}; for example, $R_Z(2\pi)=-I$ is equal to $I$ only up to global phase, despite its identity category. General fusion and reordering lie outside the rule domain. A different number of top-level measurements returns \texttt{failed}; any other unestablished relation returns \texttt{unsupported}, which proves neither equivalence nor inequivalence. Requests for \textsf{UP\_TO\_LOCAL} or \textsf{ENTANGLEMENT} raise an explicit error. These are executable rule checks, not a symbolic proof or a general dynamic-circuit equivalence procedure. Canonical JSON and SHA-256 identify the recorded structure, not an equivalence class.

\Needspace{5\baselineskip}
\subsection{The EmuPlat Backend Pipeline and Its Invariants}
\label{sec:six-stage}
\begin{figure}[htbp]
\centering
\includegraphics[width=\linewidth]{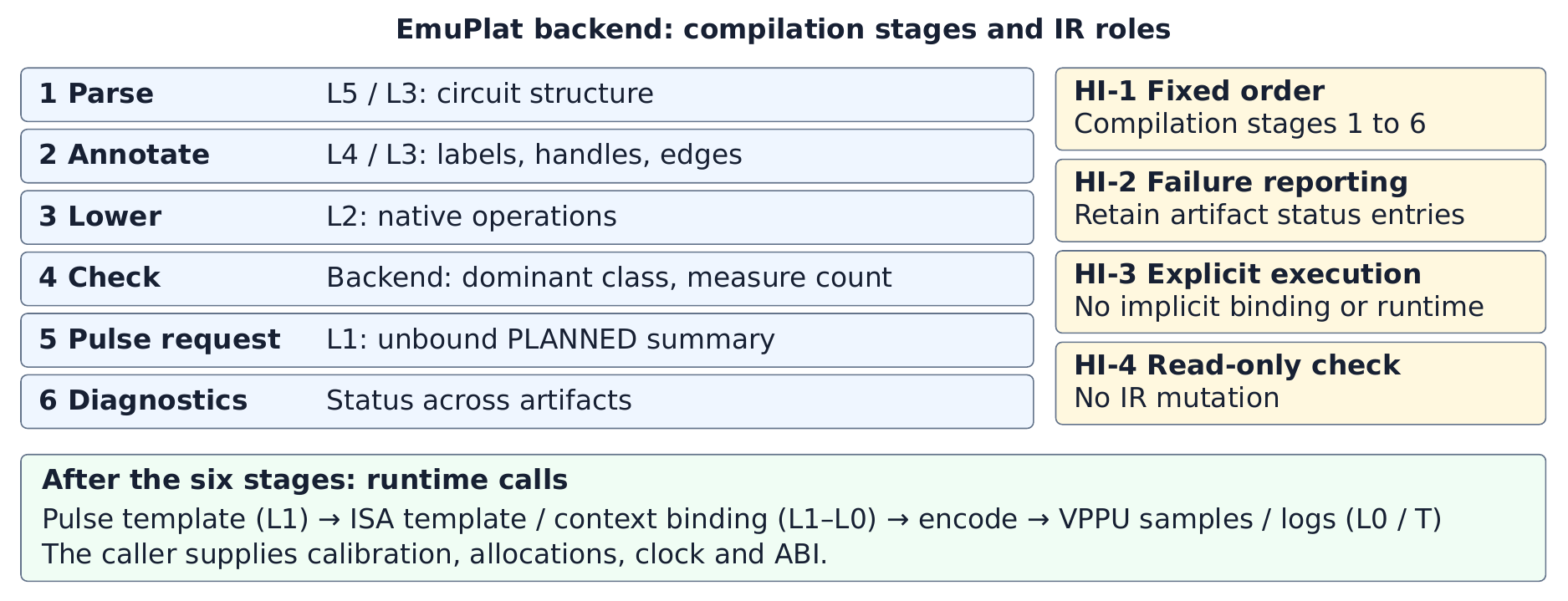}
\Description{In the EmuPlat backend, six compilation stages map to overlapping IR roles: parse L5/L3, annotate L4/L3, lower L2, check L4/L2, pulse request L1, collect diagnostics across layers. Template construction, device binding, and VPPU execution follow as separate calls. Four invariants describe the compiler API.}
\caption{EmuPlat backend: compilation stages, IR roles, and runtime calls.}
\label{fig:pipeline}
\end{figure}

The adapted canonical QASM (\S\ref{sec:m3-walkthrough}) enters EmuPlat, which reparses the source and uses its own annotation and checking code. Its six stages span several architectural roles: parsing (L5/L3), annotation (L4/L3), native-gate lowering (L2), an aggregate check (L4/L2), an unbound pulse request (L1), and diagnostics. Figure~\ref{fig:pipeline} shows this mapping. This path does not consume Qaiji-IR's validated native schedule or invoke its R1--R4 checker.

The backend aggregate baseline selects a dominant category from its input annotations, excluding measurement when other operations occur and taking the first present category in the order unitary, permutation, phase, identity, and classical control. Measurement-only input receives the measurement category. Different measurement counts produce \texttt{failed}; a category change produces \texttt{warning}; agreement produces \texttt{passed/EXACT}. Here the backend label \textsf{EXACT} records aggregate agreement, not the operator relation checked by Qaiji-IR. Its \texttt{measurement\_map} argument is unused; gate order, angles, operands, classical labels, and general predicates are absent from this comparison.

The EmuPlat API keeps four invariants: fixed stage order (HI-1), retention of status fields after failure (HI-2), no implicit hardware access or materialization (HI-3), and read-only checking (HI-4). Its default \texttt{compile\_qasm()} returns an unbound \textsf{PLANNED} pulse summary. Callers construct templates, supply allocation, clock, ABI, and calibration contexts, then invoke execution separately.

\subsection{Backend Integration Modes}
\label{sec:state-machine-and-innov}
EmuPlat's \texttt{semantic\_ir} option controls its annotation path: \textsf{off} omits semantic fields; \textsf{handle\_only} returns available annotations; \textsf{required} fails if the producer is unavailable. Backend import and fresh-process tests check separation from ISA types. These are backend configuration modes, not options of the Qaiji-IR frontend. \S\ref{sec:adversarial} compares the two checking mechanisms on identical circuit descriptions.

\subsection{Native Scheduling and Validation}
\label{sec:native-validation}
Qaiji-IR L2 expands supported gates into a CZ basis under caller-supplied durations and resources. Scheduling is deterministic and serial. The schedule retains source references, decomposition groups, measurement events, and conditional regions. A separate projection revalidates these records before emitting neutral operation labels; it supplies neither ALU/jump instructions nor a feedback runtime.

Separate routines check source correspondence and rules, compare each local native unitary product with its ideal source operator after accounting for an independently computed global phase, and check resources and timing. Matrix checks use $2\times2$ or $4\times4$ operators with elementwise tolerance $10^{-10}+10^{-8}|u_{ij}|$; measurement receives \texttt{not\_applicable}. Every bit read by a condition must reference its latest preceding measurement, and the region must start no earlier than configured result readiness. JSON decoding alone does not validate a schedule; validation requires the original circuit and configuration. No sampling, hardware ISA, waveform generation, or physical-fidelity estimate is supplied. Connecting these validated schedules to the EmuPlat execution path remains further work.

\section{Implementation of Measurement-Conditioned Control}
\label{sec:impl}
\subsection{Typed Frontend and Syntax Adaptation}
\label{sec:m3-walkthrough}
The Qaiji-IR frontend at the main commit specified in \S\ref{sec:layers} produces immutable Gate, Measure, and Conditional nodes and canonical OpenQASM~3. Parsing, serializing, and reparsing Listing~\ref{lst:reset} leaves the canonical text unchanged. The codec accepts both whole-register equality and a one-bit indexed form, but serialization emits whole-register equality. EmuPlat requires the indexed form. The one-bit adapter therefore reconciles canonical output with backend input; it is not a workaround for a missing indexed parser.

\Needspace{11\baselineskip}
\begin{lstlisting}[style=pythonsnippet,language={},basicstyle=\ttfamily\footnotesize,
 caption={Single-qubit active reset. The adapter changes the condition operand from c to c{[0]}.},label={lst:reset}]
OPENQASM 3.0;
include "stdgates.inc";
qubit[1] q;
bit[1] c;
x q[0];
c[0] = measure q[0];
if (c == 1) { x q[0]; }
\end{lstlisting}

The adapter accepts an explicitly declared one-bit register and inserts \texttt{[0]} in its equality operand. It preserves the comparison constant and conditional gate body, checks the before/after ASTs and binary truth table, and rejects wider registers or other condition forms. The experiment driver applies a separate input check: measured bit 0, \texttt{==1}, and integer outcomes 0 or 1. Thus the adapter can preserve \texttt{==0}, while the driver rejects it for the complete execution example. Both helpers belong to the experiment driver, not to Qaiji-IR or EmuPlat.

\subsection{Semantic Annotation and Preservation}
\label{sec:m1-coreleaf}
The source asks for an initial $X$, a measurement, and a conditional $X$. Qaiji-IR produces four annotations: the initial gate, measurement, conditional node, and body gate. Their categories are permutation, measurement, classical control, and permutation. Its summary records \texttt{m0} for \texttt{c[0]} and edge \texttt{m0}$\to$\texttt{cond2}, where 2 is the conditional's top-level index. This is a declared dependency; readiness is checked separately at L2.

EmuPlat reparses the adapted QASM and produces three backend annotations, its own \texttt{m0}$\to$\texttt{cond0} edge, and the condition model \textsf{REGISTER\_ALU}. Its initial $X$ lowers to two \texttt{RX90} operations. The backend aggregate check returns \texttt{warning} because the dominant category changes from permutation to unitary, and compilation continues. This is not a Qaiji-IR R1--R4 verdict, and a category warning alone does not determine the relation between ideal operators.

\subsection{Pulse and ISA Materialization}
\label{sec:m2-pipeline}
After EmuPlat compilation, callers build pulse and QubiC-like ISA templates and bind registers, envelopes, frequencies, channels, clock, ABI, and calibration identifiers. The executable bridge emits the instructions in Table~\ref{tab:trace}. These runs use placeholder pulse parameters: unit amplitude, four-cycle envelopes, and ordinal command times. The VPPU executes them with its default waveform facilities. Calibrated pulse behavior cannot be inferred from this instruction-execution test.

\subsection{VPPU Execution}
\label{sec:m4-deviceisa}
The encoded program runs with externally supplied outcomes 0 and 1. Table~\ref{tab:trace} shows the instructions and PC paths; the traces also record each opcode, feedback slot, ALU result, and branch target. Because the initial $X$ also produces drive samples, PC~6 identifies the conditional drive. VPPU triggers inject the configured feedback value. Initial drive triggers can latch it before the readout trigger, so these runs do not test the order of readout and branch.

\begin{table}[htbp]

\caption{Generated instructions and observed paths for Listing~\ref{lst:reset}. The feedback slot index is 0; PC~7 exits the program.}
\label{tab:trace}
\small
\begin{tabularx}{\linewidth}{@{}l l >{\raggedright\arraybackslash}X@{}}
\toprule
PC & Decoded opcode & Fields / observed effect \\
\midrule
0, 1 & \texttt{PULSE\_WRITE\_TRIG} & Initial drive (two native pulses), channel 0 \\
2 & \texttt{PULSE\_WRITE\_TRIG} & Readout, pulse register/channel 1 \\
3 & \texttt{ALU\_FPROC\_I} & EQ(slot, 1); destination R0 discards writes \\
4 & \texttt{JUMP\_FPROC\_I} & Raw slot $\ne0$ targets PC~6 \\
5 & \texttt{JUMP\_I} & Target PC~7 skips conditional drive \\
6 & \texttt{PULSE\_WRITE} & Conditional drive, channel 0 \\
\midrule
\multicolumn{2}{@{}l}{Slot value 0: PCs 0,1,2,3,4,5} & ALU output 0; drive at PC~6 skipped \\
\multicolumn{2}{@{}l}{Slot value 1: PCs 0,1,2,3,4,6} & ALU output 1; drive at PC~6 executed \\
\bottomrule
\end{tabularx}
\end{table}

The branch reads the raw feedback slot; the EQ result is written to R0, which is hardwired to zero. For binary values, the observed behavior agrees with \texttt{c[0]==1}. Other bit indices and predicates are rejected or execute incorrectly (\S\ref{sec:feedback-matrix}). The feedback values are injected; no simulated measurement collapses a quantum state in these runs. A calculation of reset would have to apply measurement back-action and the conditional drive to the same state trajectory.

\Needspace{9\baselineskip}
\section{Evaluation of Contract and Execution Semantics}
\label{sec:eval}
\subsection{Aggregate Checks and Reference Semantics}
\label{sec:adversarial}
I use a NumPy reference calculation for circuits of at most two qubits. It constructs complete ordered unitary matrices and compares strict equality and equality up to global phase, using absolute tolerance $10^{-10}$ and relative tolerance zero. For measurement, it compares each labeled quantum operation through its superoperator $\sum_k\overline{K_{y,k}}\otimes K_{y,k}$, retaining the classical register label $y$. The labeled map retains measurement back-action and its classical output. This comparison covers all quantum input states; global Kraus phase cancels within each labeled operation. Five analytic controls check identity products, the $R_Z(2\pi)=-I$ boundary, noncommutation, and label sensitivity. The calculation uses ideal circuit operators and excludes pulse waveforms.

Table~\ref{tab:adversarial} compares Qaiji-IR with the EmuPlat aggregate baseline and the ideal reference on the same nine circuit pairs. All pairs pass the baseline's construction and operand-bound checks. Its aggregate test detects measurement loss and a dominant-category change, but passes four behavior changes and the strict-phase difference between $R_Z(2\pi)$ and $I$. The Qaiji-IR results are obtained by calling its circuit-to-circuit checker directly; they do not come from the backend compile call.

\begin{table}[htbp]

\caption{Checker comparison on identical inputs. Aggregate: EmuPlat backend; Qaiji-IR: commit \texttt{4ef6f1a}. E/P: \textsf{EXACT}/\textsf{UP\_TO\_PHASE}. S/G: strict/global-phase reference equality, label preserving for measurement.}
\label{tab:adversarial}
\footnotesize
\begin{tabularx}{\linewidth}{@{}>{\raggedright\arraybackslash}X l l cc@{}}
\toprule
Original $\to$ candidate & Aggregate & Qaiji-IR & S & G \\
\midrule
Identical $H;X$ & passed/E & passed/E & yes & yes \\
$R_X(.2);R_X(.3)\to R_X(.5)$ & passed/E & unsupported & yes & yes \\
$H;M\to H$ (measurement lost) & failed & failed & no & no \\
$H\to X$ (dominant category changes) & warning & unsupported & no & no \\
$H;X\to X;H$ & passed/E & unsupported & no & no \\
$R_X(.3)\to R_X(.7)$ & passed/E & unsupported & no & no \\
$R_X^{q_0}(.3)\to R_X^{q_1}(.3)$ & passed/E & unsupported & no & no \\
$H;M_{q_0\to c_0}\to H;M_{q_0\to c_1}$ & passed/E & unsupported & no & no \\
$R_Z(2\pi)\to I$ & passed/E & passed/P & no & yes \\
\bottomrule
\end{tabularx}
\end{table}

Changing the measurement destination changes the labeled quantum instrument while preserving both aggregates. The final row separates a registry category from matrix equality: $R_Z(2\pi)$ receives an identity tag but differs from $I$ by global phase. Conversely, the backend reset example in \S\ref{sec:m1-coreleaf} produces a category warning when $X$ decomposes into two \texttt{RX90} operations, whose product equals $X$ up to global phase. Category agreement and disagreement both require an operation-level check to determine equivalence.

Qaiji-IR also fails the lost measurement. It abstains on $H\to X$ and on the four behavior-changing pairs accepted by the aggregate baseline. For $R_Z(2\pi)\to I$, it reports \textsf{UP\_TO\_PHASE}; requesting \textsf{EXACT} instead returns \texttt{unsupported}. It also abstains on valid rotation fusion, demonstrating incomplete coverage rather than a general equivalence decision. In a separate L2 test on the same reset circuit, synthetic 4-ns native pulses, 10-ns readout, and 6-ns result latency place the condition at 24~ns and completion at 32~ns. Validation accepts the JSON-restored schedule and rejects independent mutations to its source reference, declared phase, pulse axis, and duration, plus a condition without preceding measurement. These are static configuration checks, not measured device latency; this schedule is not the VPPU input.

\Needspace{5\baselineskip}
\subsection{Predicate Support}
\label{sec:functional-indicators}
\label{sec:feedback-matrix}
Table~\ref{tab:feedback} follows seven conditions through parsing, classification, materialization, and execution in EmuPlat. All seven parse as OpenQASM ASTs. A separate ten-case test of Qaiji-IR accepts whole-register equality and multi-gate bodies, plus a one-bit indexed equality; it rejects indexing a wider register, inequality, compound predicates, and else. Canonical serialization still emits whole-register equality, explaining the adapter in \S\ref{sec:m3-walkthrough}. Table~\ref{tab:feedback} reports backend behavior, not the Qaiji-IR codec or its L2 condition validator.

\begin{table}[htbp]

\caption{EmuPlat predicate matrix. A: AST parses; C: feedback scan and L4 classify; M: device ISA materializes; E: encodes and runs. Dash means not reached.}
\label{tab:feedback}
\small
\begin{tabularx}{\linewidth}{@{}l cccc >{\raggedright\arraybackslash}X@{}}
\toprule
Condition/body & A & C & M & E & Result \\
\midrule
\texttt{c[0]==1}; one $X$ & yes & yes & yes & yes & Matches source for injected 0/1 \\
\texttt{c[0]==0}; one $X$ & yes & yes & yes & yes & Executes inverse condition for 0/1 \\
\texttt{c[1]==1}; one $X$ & yes & yes & yes & -- & Bridge refuses nonzero func\_id \\
\texttt{c[0]!=0} & yes & no & -- & -- & Dynamic-feedback scan rejects \\
Compound \texttt{\&\&} & yes & no & -- & -- & Dynamic-feedback scan rejects \\
Equality with else & yes & no & -- & -- & Dynamic-feedback scan rejects \\
Equality; two-gate body & yes & no & -- & -- & Dynamic-feedback scan rejects \\
\bottomrule
\end{tabularx}
\end{table}

For \texttt{==0}, lowering still compares to 1 and execution tests whether the raw slot is nonzero: slot value 0 skips the drive, and value 1 executes it. This reverses the source condition. With \texttt{==1}, the injected integer 2, outside the range of a single-bit outcome, produces ALU output 0 but still executes the drive. The matrix runs bypass the experiment driver's binary input check to exercise these cases. In the complete active-reset example, the driver accepts only \texttt{==1} and binary outcomes.

\Needspace{5\baselineskip}
\section{Related Work}
\label{sec:related}
\label{sec:related-existing-irs}
Table~\ref{tab:comparison} relates Qaiji-IR and the backend experiment to MLIR, QIR, Catalyst, and qe-compiler. These systems provide metadata, verification, or lowering mechanisms on which a hybrid compiler can build. Qaiji-IR specifies operation structure, bounded equivalence rules, and native validation; EmuPlat supplies the external pulse/ISA path. The comparison concerns these mechanisms, not a general claim of superiority.

\begin{table}[htbp]

\caption{Compiler mechanisms related to Qaiji-IR and the backend experiment.}
\label{tab:comparison}
\scriptsize
\begin{tabularx}{\linewidth}{@{}l>{\raggedright\arraybackslash}X>{\raggedright\arraybackslash}X@{}}
\toprule
System & Documented mechanism & Relation to this prototype \\
\midrule
MLIR~\cite{mlir,mlir-langref} & Operations, typed SSA values, attributes, traits/interfaces, and verification constraints; dialect conversion & Attributes can represent L4 records; Qaiji-IR specifies position-aligned rules and native validation. \\
QIR~\cite{qir2021} & LLVM representation, runtime/QIS functions, and target capability profiles & Possible lowering target, not evaluated here. Qaiji-IR keeps operation structure and equivalence reports. \\
Catalyst~\cite{catalyst} & MLIR quantum dialect, hybrid classical/quantum optimization, LLVM/QIR lowering, and runtime & Both address hybrid compilation; the example here follows one source program into a conditional pulse/ISA program. \\
qe-compiler~\cite{ibm-qe-pulse,qe-source} & OQ3, QUIR, Pulse, and QCS dialects; localization passes and target framework & Also spans circuit and pulse levels. Qaiji-IR checks circuit preservation and native schedules; EmuPlat supplies the pulse path. \\
\bottomrule
\end{tabularx}
\end{table}

OpenQASM~3 provides classical control, timing, and pulse calibration constructs~\cite{openqasm3}; the prototype consumes a restricted subset. CUDA-Q~\cite{cudaq} and QLLVM~\cite{qllvm} provide other quantum-classical compilation infrastructures. Qiskit Pulse~\cite{qiskit-pulse} and MQSS Pulse~\cite{mqss-pulse} motivate pulse interfaces across runtime boundaries. QubiC~2.0 supplies a hardware feed-forward control stack~\cite{qubic2}. The VPPU example here uses injected feedback values to test the corresponding instruction path.

L4 associates operation categories with measurement dependencies and explicit operation structure. In Table~\ref{tab:adversarial}, the EmuPlat aggregate baseline adds two diagnostics beyond construction and operand-bound checks. Qaiji-IR's rule checker abstains on four further behavior changes and distinguishes the full-turn phase relation, but leaves valid fusion undecided. Its added information is therefore specific and bounded; these cases do not establish a general cross-layer preservation proof.

\label{sec:related-quadrant}
\label{sec:related-pulse-isa}
\noindent\textbf{Limitations.}\label{sec:related-caveats}
The execution example combines the evaluated Qaiji-IR frontend with EmuPlat, bit-0 equality to one, and injected binary outcomes. The backend does not consume the validated Qaiji-IR schedule, so native validation does not certify pulse/ISA execution. Rule coverage, local matrices, and configured serial timing do not establish general dynamic-circuit equivalence or calibrated pulse correctness. L1/L0/P/T are outside the Qaiji-IR public API. Backend dynamics models and solver APIs exist, but these experiments do not continue a measured state through conditional evolution. Logs and certificate metadata are not automatic preservation proofs. Integrated traces and physical feedback remain further work.

\Needspace{5\baselineskip}
\section{Conclusion and Future Work}
\label{sec:conclusion}
Qaiji-IR organizes the relation between ideal quantum operations, measurement records, and device controls through a five-axis contract. L4 records operation categories, equivalence labels, optimization hints, and measurement-to-condition dependencies, while L2 records native schedules under explicit timing and resource configurations.

All Qaiji-IR results refer to one evaluated main commit. Its L4 rules decide a bounded set of operator relations, while its L2 routines separately check source correspondence, decomposition, local matrices, and configured timing. Circuit-pair tests compare those decisions with ideal unitary action and labeled measurement operations. On nine pairs, the rules pass none of the four behavior changes accepted by an aggregate baseline, but leave valid rotation fusion undecided. The same active-reset source passes through its frontend and a syntax adapter to EmuPlat pulse/ISA generation and VPPU execution. Backend predicate tests expose conditions that compile but execute incorrectly.

Further research should extend this implementation's transformation coverage and connect its validated native schedules to pulse/ISA lowering. Backend jumps must also use the predicate result instead of the raw feedback slot. Connecting measurement collapse to conditional continuation in one dynamics trajectory is the next step toward physical reset and measurements of fidelity and latency. Integrated P/T traces and formal cross-layer refinement remain open tasks.


\begin{credits}
  \subsubsection{\ackname}
    The author thanks Prof.~Lu Jun, a member of the Chinese Academy of Engineering, for inspiring the concept of cross-domain logic on which this work is based.

  \subsubsection{\discintname}
    The author has no competing interests to declare that are relevant to the content of this article.
\end{credits}

\bibliographystyle{splncs04}
\Needspace{5\baselineskip}
\bibliography{references}

\begin{thebibliography}{10}
\providecommand{\url}[1]{\texttt{#1}}
\providecommand{\urlprefix}{URL }
\providecommand{\doi}[1]{https://doi.org/#1}

\bibitem{qiskit-pulse}
Alexander, T., Kanazawa, N., Egger, D.J., Capelluto, L., Wood, C.J.,
  Javadi-Abhari, A., McKay, D.C.: {Qiskit Pulse}: Programming quantum computers
  through the cloud with pulses. Quantum Science and Technology  \textbf{5}(4),
   044006 (2020). \doi{10.1088/2058-9565/aba404}

\bibitem{openqasm3}
Cross, A., Javadi-Abhari, A., Alexander, T., de~Beaudrap, N., Bishop, L.S.,
  Heidel, S., Ryan, C.A., Sivarajah, P., Smolin, J., Gambetta, J.M., Johnson,
  B.R.: {OpenQASM} 3: A broader and deeper quantum assembly language. ACM
  Transactions on Quantum Computing  \textbf{3}(3),  1--50 (2022).
  \doi{10.1145/3505636}

\bibitem{mqss-pulse}
Echavarria, J., Farooqi, M.N., Devra, A., Lujan, S., Van~Damme, L., Ahmed, H.,
  Letras, M., Kaya, E., Vetter, A., Werninghaus, M., Knudsen, M., Rohde, F.,
  Frisch, A., Mansfield, E., Davletkaliyev, R., Kukushkin, V.,
  F\"{a}rkkil\"{a}, N., M\"{a}ntyl\"{a}, J., Pomplun, N., Sp\"{o}rl, A.,
  Burgholzer, L., Stade, Y., Wille, R., Schulz, L.B., Schulz, M.: Tackling the
  challenges of adding pulse-level support to a heterogeneous {HPCQC} software
  stack: {MQSS} pulse. In: Proceedings of the SC '25 Workshops of the
  International Conference for High Performance Computing, Networking, Storage
  and Analysis. pp. 1868--1878. ACM (2025). \doi{10.1145/3731599.3767552}

\bibitem{ibm-qe-pulse}
Healy, M.B., Jokar, R., Thomas, S., Pascuzzi, V.R., Barton, K., Alexander,
  T.A., Elkabetz, R., Donovan, B.C., Horii, H., Hillenbrand, M.: Design and
  architecture of the {IBM} {Q}uantum {E}ngine {C}ompiler. In: 2024 IEEE
  International Conference on Quantum Computing and Engineering (QCE). pp.
  866--872. IEEE (2024). \doi{10.1109/QCE60285.2024.00106}

\bibitem{mlir}
Lattner, C., Amini, M., Bondhugula, U., Cohen, A., Davis, A., Pienaar, J.,
  Riddle, R., Shpeisman, T., Vasilache, N., Zinenko, O.: {MLIR}: Scaling
  compiler infrastructure for domain specific computation. In: Proceedings of
  the 19th IEEE/ACM International Symposium on Code Generation and Optimization
  (CGO). pp. 2--14. IEEE (2021)

\bibitem{mlir-langref}
{LLVM Project}: {MLIR} language reference.
  \url{https://mlir.llvm.org/docs/LangRef/}, accessed: 2026-10-17

\bibitem{cudaq}
{NVIDIA Corporation}: {CUDA-Q}: A heterogeneous quantum-classical programming
  model in modern {C++}. \url{https://developer.nvidia.com/cuda-q}; source:
  \url{https://github.com/NVIDIA/cuda-quantum} (2024), accessed: 2026-09-19

\bibitem{qllvm}
{QCFlow Consortium}: {QLLVM}: Quantum {LLVM} compiler infrastructure.
  \url{https://github.com/QCFlow/QLLVM} (2026), accessed: 2026-09-19

\bibitem{qe-source}
{qe-compiler contributors}: {qe-compiler}: Compiler framework and dialect
  overview. \url{https://github.com/openqasm/qe-compiler}, accessed: 2026-10-17

\bibitem{qir2021}
{QIR Alliance}: Quantum intermediate representation specification.
  \url{https://github.com/qir-alliance/qir-spec} (2021), accessed: 2026-10-17

\bibitem{song1637}
Song, Y.: T'ien-Kung K'ai-Wu: Chinese Technology in the Seventeenth Century.
  Pennsylvania State University Press (1966), english translation of the work
  first published in 1637

\bibitem{catalyst}
Xanadu: {Catalyst}: Quantum just-in-time compiler.
  \url{https://github.com/PennyLaneAI/catalyst} (2024), official compiler
  architecture overview. Accessed: 2026-10-17

\bibitem{qubic2}
Xu, Y., Huang, G., Fruitwala, N., Rajagopala, A., Naik, R.K., Nowrouzi, K.,
  Santiago, D.I., Siddiqi, I.: {QubiC} 2.0: An extensible open-source qubit
  control system capable of mid-circuit measurement and feed-forward (2023).
  \doi{10.48550/arXiv.2309.10333}

\bibitem{qaiji-ir}
Ye, J.: {Qaiji-IR}: Quantum-classical intermediate representation.
  \url{https://github.com/yjmaxpayne/Qaiji-IR}, main branch, commit
  \href{https://github.com/yjmaxpayne/Qaiji-IR/commit/4ef6f1a07a9e03ca09fff13f7843e0404958f228}{\texttt{4ef6f1a}}.
  Verified: 2026-10-18

\bibitem{emuplat}
Ye, J., Khoo, J.Y.: {EmuPlat}: A framework-agnostic platform for quantum
  hardware emulation with validated transpiler-to-pulse pipeline. In:
  Proceedings of the Supercomputing Asia and International Conference on High
  Performance Computing in Asia Pacific Region (SCA/HPCAsia 2026). pp. 10--18.
  ACM (2026). \doi{10.1145/3773656.3773674}, osaka, Japan

\end{thebibliography}

\end{document}